\documentstyle[prl,aps,amsfonts,preprint,epsfig,floats]{revtex}
 
\tighten

\newcommand{\be}{\begin{eqnarray}}
\newcommand{\ee}{\end{eqnarray}}
\newcommand{\dslash}{\partial \hskip -0.5em /}
\newcommand{\ID}{\mbox{{\sf 1}\hspace{-0.55mm}\rule{0.04em}{1.53ex}}}

\begin{document}

\title{A Heavy Fermion Can Create a Soliton: A 1+1~Dimensional Example}
\author{E.~Farhi\footnote[0]{e-mail:  farhi@mit.edu,
graham@pierre.mit.edu, jaffe@mit.edu, weigel@ctp.mit.edu}$^{\rm a}$,
N.~Graham$^{\rm a,b}$,
R.~L.~Jaffe$^{\rm a}$, and
H.~Weigel{}\footnote{Heisenberg Fellow}$^{\rm a}$}

\address{{~}\\$^{\rm a}$Center for Theoretical Physics, Laboratory for
	Nuclear Science
	and Department of Physics \\
	Massachusetts Institute of Technology,
	Cambridge, Massachusetts 02139 \\
	and \\
	$^{\rm b}$ Dragon Systems, Inc.  Newton, MA 02460\\
	{~}
	{\rm MIT-CTP\#2937 \qquad hep-th/9912283}}

\maketitle

\begin{abstract}
We show that quantum effects can stabilize a soliton in a model with
no soliton at the classical level.  The model has a scalar field
chirally coupled to a fermion in 1+1~dimensions.  We use a formalism
that allows us to calculate the exact one loop fermion contribution to
the effective energy for a spatially varying scalar background.  This energy
includes the contribution from counterterms  fixed in the perturbative
sector of the theory.  The resulting energy is therefore finite and
unambiguous.  A variational search then yields a fermion number one
configuration whose energy is below that of a single free fermion.
\end{abstract}

\section{Introduction}

Scalar field theories can contain spatially varying (but time independent) 
configurations that are local minima of the classical energy. These 
solitons are found as solutions to the non--linear classical equations of
motion.  Sometimes a topological conservation law can be used to  show that
the soliton is absolutely stable  because it is the lowest energy
configuration with a given value of a conserved topological charge.  When
quantum effects are taken into account, the classical description must be
re--examined. Now the spatially varying soliton configuration should
minimize the ``effective energy'' which takes into account classical and
quantum effects\footnote{By effective energy we mean the effective action
per unit time; the term ``effective potential'' is reserved for spatially
constant configurations.}. Since the effective energy for general
configurations is difficult to compute, quantum effects are typically
computed as approximate corrections to the classical soliton.  In a
non--renormalizable theory, these corrections are cutoff dependent and the
model must be redefined to include the cutoff prescription.  The hope is
that the energy of the soliton is slightly altered by quantum effects but
its qualitative  features  remain. 

In this Letter we give an example of a quantum soliton that is not present
in the classical theory alone.  We examine a renormalizable model in $1+1$
dimensions where a scalar field is Yukawa coupled to a fermion. Fermion
number is conserved. The classical energy is minimized when the scalar
field has a constant value $v$.  There are no classical solitons.  The
fermion gets a mass $m=Gv$ through the Yukawa coupling. We calculate exactly
the fermion's properly renormalized one loop contribution to the scalar
field effective energy. By ``exactly'' we mean to all orders in the
derivative expansion, which is crucial since we consider configurations 
varying on the scale $1/m$. We then show that for certain choices of model 
parameters --- in particular with $G$ large --- we can exhibit a field 
configuration that carries fermion number and has energy below $m$.  It
cannot decay by emitting a free fermion.  We search for the lowest energy
configuration carrying fermion number using a few parameter variational
{\it ansatz}.  The soliton, which is the actual lowest energy configuration
with fermion number one, is presumably not far from our variational
minimum and has strictly lower energy.  The soliton therefore has energy
less than $m$ and is absolutely stable.

The idea that a heavy fermion can create as soliton is not new and has been
explored previously \cite{prevwork,Farhi}.  What we are adding to the
discussion is the ability to exactly calculate the renormalized
fermionic one loop effective energy for any spatially varying meson
background, which is essential for demonstrating stability at the
quantum level.  Since we are working in a renormalizable theory, the
counterterms in the Lagrangian cancel the cutoff dependent part of the
sum over zero--point energies in the explicit evaluation of the
effective energy, leaving a finite result.  This result is unambiguous
because we are able to fix the counterterms in the perturbative sector
of the theory.  Furthermore, we can choose model parameters to
justify neglecting the one loop boson contributions and all higher loop
contributions. Thus we conclude that in this $1+1$ dimensional model,
a heavy fermion can create a soliton.

\section{The Model}

The model we consider has a two--component meson field 
$\vec{\phi}=(\phi_1,\phi_2)$ coupled equally to $N_F$ fermions. We 
suppress the fermion flavor label but will keep track of the factor 
$N_F$ as necessary.
The Lagrangian is ${\cal L}={\cal L}_B+{\cal L}_F$ with
\be
{\cal L}_B=\frac{1}{2}\, \partial_\mu\vec{\phi}\cdot\partial^\mu\vec{\phi}
-V(\vec{\phi})\, .
\label{lagb}
\ee
where
\be
V(\vec{\phi})=\frac{\lambda}{8}
\left[\vec{\phi}\cdot\vec{\phi}
-v^2+\frac{2\alpha v^2}{\lambda}\right]^2
-\frac{\lambda}{2}\left(\frac{\alpha v^2}{\lambda}\right)^2
-\alpha v^3\left(\phi_1-v\right)\,
\label{lag2}
\ee
and
\be
{\cal L}_F=\frac{1}{2}\left(i\left[\bar{\Psi},\dslash\Psi\right]
- G\left(\left[\bar{\Psi},\Psi\right]\phi_1
+i\left[\bar{\Psi},\gamma_5\Psi\right]\phi_2\right) \right)\, .
\label{lagf}
\ee
(The reason for the commutators in eq.~(\ref{lagf}) will be explained
later.) Note that with $\alpha$ set to zero, the theory has a global 
$U(1)$ invariance
\be
\phi_1+i\phi_2\to e^{i\varphi} 
\left(\phi_1+i\phi_2\right)
\quad {\rm and} \quad
\Psi\to e^{-i\varphi\gamma_5/2}\Psi\, .
\label{u1inv}
\ee
Na\"\i vely, one would imagine that spontaneous symmetry breaking
occurs with $\alpha=0$.  Then we could pick a classical vacuum, say
$\vec{\phi}_{\rm cl}=(v,0)$, and expand the theory about this point.
But in $1+1$ dimensions, the massless mode that corresponds to motion
along the chiral circle, $\vec{\phi}\cdot\vec{\phi}=v^2$, gives rise
to infra--red singularities and there is no spontaneous symmetry breaking
\cite{Coleman1d}.  By introducing $\alpha\ne0$ we have
tilted the potential to eliminate the massless mode. For $\alpha$ large
enough it is legitimate to expand about $\vec\phi_{\rm cl}$. There are two
massive bosons, which we call $\sigma$ and $\pi$, with
$m_\sigma^2=\left(\lambda+\alpha\right)v^2$ and  $m_\pi^2=\alpha
v^2$. The classical bosonic theory governed by ${\cal L}_B$ has no
classical soliton.

The fermions get mass through their Yukawa coupling to $\vec{\phi}$. In the
perturbative vacuum, expanding about $\vec\phi_{\rm cl}$, the fermion  has
mass $m=Gv$.  One could imagine that various distortions of $\vec\phi$ 
would affect the fermion spectrum.  For example, one could keep $\phi_2=0$ 
and let $\phi_1\to\phi_1(x)$ with  $\lim_{x\to\pm\infty}\phi_1(x)=v$, but 
$\phi_1(x)<v$ over some region in $x$ of order $w$. Alternatively, one
could keep  $\vec\phi\cdot\vec\phi=v^2$, but let 
$\vec\phi=v(\cos\theta(x),\sin\theta(x))$, where $\theta(x)
\to 0$ as $x\to-\infty$ and $\theta(x)\to 2\pi$ as $x\to+\infty$.  Again 
the deviation of $\vec\phi$ from $\vec\phi_{\rm cl}$ occurs in a region of
order $w$.  In both cases, if $w$ is of order $1/m$, then there are
bound state solutions of the single--particle Dirac equation associated with
eq.~(\ref{lagf}) that have binding energies of order $m$, so that a
fermion bound to the $\vec{\phi}$ field has an energy below $m$.
(Because of its topological properties, the latter configuration is
especially efficient at binding a fermion \cite{longpaper}.)  However,
there is an energy cost from the gradient and potential terms.  Still,
considering just the single bound fermion and the classical scalar
energy, we might expect a total energy below $m$ for $G$ large enough.

Of course, $\Psi$ describes a quantum field and any distortion of the
background $\vec{\phi}(x)$ away from $\vec{\phi}_{\rm cl}$ will cause shifts
in the zero--point energies of the fermion fluctuations.  To form a
self--consistent approximation, we must compute the effect of these shifts
as well, since they are of the same order in $\hbar$ as the bound state
contribution.  In general the sum over zero--point energies diverges.  In
order to proceed we must regularize and renormalize the calculation.  We
are working in a renormalizable field theory so we know that the counterterms
that are implicit in ${\cal L}_B$ will cancel these divergences.  We want
to compare the energy of non--trivial configurations with the perturbative
spectrum of the model, therefore we fix the counterterms by standard
renormalization conditions on the Green's functions.  The Green's functions
are evaluated perturbatively so the counterterms have an expansion in
Feynman diagrams.

Regularization and renormalization of the sum over zero--point energies has
been problematic in the past.  We work in the continuum, where
the sum is replaced by an integral over scattering phase shifts.  In
Ref.~\cite{longpaper} we show that it is possible to analytically continue
this integral to $d$ spatial dimensions, where it converges.  Then we
are able to identify potential divergences with low orders in the Born
expansion for the phase shifts, and, in turn, with specific Feynman
diagrams.  We subtract the low order Born terms from the integral,
which then remains finite when analytically continued back to $d=1$.
We then add back in the corresponding Feynman diagrams, which combine
with the counterterms in the usual way to yield a finite and
unambiguous result in $d=1$.  In the next section we show how we
evaluate this contribution to the energy of a static configuration.

\section{The One Fermion Loop Effective Energy}

We have written eq.~(\ref{lagf}) as a commutator to ensure that the Lagrangian
is invariant under the charge conjugation operation
$\Psi\to{\mathbb{C}}\Psi^\ast$ and  $(\phi_1,\phi_2)\to(\phi_1,-\phi_2)$. 
As a result, the vacuum energy gets contributions from both the
positive and negative energy eigenvalues of the single particle Dirac
Hamiltonian
\be
H[\vec{\phi}\,] = i\sigma_1 \frac{d}{dx}
+G\left(\sigma_2\phi_1+\sigma_3\phi_2\right)\, .
\label{dirham}
\ee
Here we are using a Majorana representation of the Dirac matrices,
$\gamma^0= \sigma_2$, $\gamma^1=i\sigma_3$, and $\gamma_5=\sigma_1$, which
implies that ${\mathbb{C}}=\ID\,$.  For one flavor of fermions the energy
of the lowest energy state is 
\be
E_{\rm vac}=\frac{1}{2}\left\{-\sum_{\omega_n>0} \omega_n
+\sum_{\omega_n<0} \omega_n \right\}
=-\frac{1}{2}\sum_{n}|\omega_n| \, ,
\label{energ_vac}
\ee where the $\omega$'s are the eigenvalues of $H[\vec{\phi}\,]$. (For a
charge conjugation invariant background, for each eigenvalue $\omega$ there
is an eigenvalue $-\omega$, so the two sums in eq.~(\ref{energ_vac}) are the
same, and $E_{\rm vac}$ reduces to the sum over the Dirac sea, $E_{\rm vac} =
\sum_{\omega_n < 0} \omega_n$.)  We will restrict our attention to
background fields that obey $\phi_1(x)=\phi_1(-x)$ and
$\phi_2(x)=-\phi_2(-x)$. In this case, $H[\vec{\phi}\,]$ commutes with the
parity operator $P=\sigma_2\Pi$, where $\Pi$ is the coordinate reflection
operator that transforms $x$ to $-x$.  We can thus decompose the solutions
of eq.~(\ref{energ_vac}) into separate parity channels.

For a given background, we wish to evaluate eq.~(\ref{energ_vac}) and subtract
from it what we get in the free case, $\vec{\phi}=(v,0)$.
Following Ref.~\cite{method}, we use the relationship between the change in the 
density of states and the derivative of the phase shift
\be
\rho(k)-\rho_0(k)=\frac{1}{\pi}\frac{d\delta(k)}{dk}\,
\label{statedense}
\ee
to write the change in the vacuum energy as
\be
\Delta E^F = -\frac{1}{2}\sum_l|\omega_l| -\int_0^\infty \frac{dk}{2\pi}
\omega(k) \frac{d}{dk}\delta_{F}(k) + \frac{m}{2} + E_{\rm ct}\, ,
\label{cas1}
\ee
where
\be
\delta_{F}(k)=\delta_+(\omega(k))+\delta_+(-\omega(k))
+\delta_-(\omega(k))+\delta_-(-\omega(k)) \, .
\label{totdel}
\ee
Here $\omega(k) = \sqrt{k^2+m^2}$, $\delta_\pm$ is the scattering phase
shift for the positive (negative) parity channel, the $\{\omega_l\}$
are the discrete bound state energy levels, and $E_{\rm ct}$ is the
counterterm contribution, which is fixed by renormalization conditions
discussed below.  The extra $m/2$ reflects an important subtlety in
one dimension:  in  the non--interacting case ($\delta_F(k)=0$) there
are bound states exactly at the continuum thresholds, $\omega=\pm m$,
which count as $1/2$ in the sum in eq.~(\ref{energ_vac})
\cite{Barton}.  Levinson's theorem,
\be
\delta_\pm(m) + \delta_\pm(-m) &=& \pi (n^\pm - \frac{1}{2}) \, ,
\ee
relates the phase shift at threshold to the number of bound states
$n_\pm$, with threshold bound states again counting as $1/2$.  It
allows us to rewrite eq.~(\ref{cas1}) as
\be
\Delta E^F=-\frac{1}{2}\sum_l\left(|\omega_l|-m\right)
-\int_0^\infty \frac{dk}{2\pi}\left(\omega(k)-m\right)
\frac{d}{dk}\delta_{F}(k) + E_{\rm ct} \, ,
\label{cas2}
\ee
which is convenient because it makes it clear that as we 
increase the background and a bound state appears, there are 
no discontinuities in $\Delta E^F$.

Of course, $\Delta E^{F}$ given by eq.~(\ref{cas2}) is formally infinite. 
For large $k$, the phase shifts go to zero like $1/k$  so the integral is
divergent.  To regulate this divergence and allow us to identify it
unambiguously with  specific Feynman diagrams, we have extended the method
of dimensional regularization to the density of states written in terms of
phase shifts.   The details are presented in Ref.~\cite{longpaper}.  Once
continued to $d$-dimensions, where all quantities are finite, we can
identify the leading large $k$ behavior of $\delta_{F}(k)$ with the
contribution of the first Born approximation plus the piece of the
second Born approximation related to it by chiral symmetry, which we
call $\hat \delta(k)$.  We also identify it unambiguously with the
coefficient of the Lagrangian counterterm,
$v^2-\vec\phi\cdot\vec\phi$, evaluated by standard Feynman
perturbation theory.  The renormalization conditions that fix the
counterterm in perturbation theory here translate into the statement
that in evaluating eq.~(\ref{cas2}) we should subtract $\hat
\delta(k)$ from $\delta_F(k)$.  After this subtraction the
integral can be analytically continued back to $d=1$ to give a result
that is finite and unambiguous,
\be
\Delta E^F=
-\frac{1}{2}\sum_l\left(|\omega_l|-m\right)
-\int_0^\infty \frac{dk}{2\pi}\left(\omega(k) - m\right)
\frac{d}{dk}\left(\delta_{F}(k)-\hat\delta(k)\right)\, ,
\label{cas3}
\ee
where 
\be
\hat\delta(k)=\frac{2 G^2}{k}\int_0^\infty dx
\left(v^2-\vec{\phi}(x)^2\right)\, .
\label{born}
\ee

We can solve numerically for the phase shift $\delta_{F}(k)$
for any background $\vec{\phi}(x)$.  We make use of the fact
that $P$ commutes with $H[\vec{\phi}\,]$ so that positive and negative 
parity channels are decoupled. The positive (negative) parity states
obey
\be
\psi_{\pm}(-x)=\pm \sigma_2\psi_{\pm}(x)
\label{parity1}
\ee
and therefore 
\be
\psi_{\pm}(0)\propto \pmatrix{1 \cr \pm i}\, .
\label{parity2}
\ee
Any state defined for $x\ge0$ and obeying one of the boundary
conditions in eq.~(\ref{parity2}) can be extended via
eq.~(\ref{parity1}) to the whole line, so we need only consider the
half line $x\ge0$ with the boundary conditions eq.~(\ref{parity2}).
Consider the free case, $\vec{\phi}=(v,0)$. For each energy $\omega$,
both positive and negative, the right moving eigenstate of
eq.~(\ref{dirham}) is
\be
\varphi^0_{k}(x)=\frac{1}{\omega}
\pmatrix{\omega \cr -k+im} e^{ikx} 
\label{right}
\ee
and the left moving free eigenstate is
\be
\varphi^0_{-k}(x)=\frac{1}{\omega}
\pmatrix{\omega \cr k+im} e^{-ikx} 
\label{left}
\ee
where $k=\sqrt{\omega^{2}-m^{2}}$ is positive.  For backgrounds
$\vec{\phi}(x)$ that are not everywhere equal to $\vec\phi_{\rm cl}$,
we still impose the requirement that $\vec{\phi}(x)$ approaches
$\vec\phi_{\rm cl}$ as $x$ gets large.  For these non--trivial
backgrounds we call $\varphi_{k}(x)$ the eigenstate of
eq.~(\ref{dirham}) that approaches $\varphi^0_{k}(x)$ as $x\to\infty$
and  $\varphi_{-k}(x)$ the eigenstate  that approaches
$\varphi^0_{-k}(x)$ as $x\to\infty$. Note that $\varphi^0_{k}$,
$\varphi^0_{-k}$, $\varphi_{k}$ and $\varphi_{-k}$ are not eigenstates
of $P$.  For $x\ge0$ let
\be
\psi_{\pm}(x)&=&\varphi_{-k}(x) \pm {\textstyle \frac{-ik+m}{\omega}}\, 
e^{2i\delta_{\pm}(\omega)}\varphi_{k}(x)
\label{smat}
\ee
be the parity eigenstates of $H[\vec{\phi}\,]$ with energy $\omega$.
This defines the phase shifts $\delta_\pm(\omega)$.  The $2\pi$
ambiguity in this definition is resolved by requiring that the phase
shifts be smooth and go to zero as $\omega\to\pm\infty$.  Note that in
the free case $\delta_\pm(\omega)=0$.  For any value of $\omega$, we
can solve numerically for the eigenstates of eq.~(\ref{dirham}) in
both parity channels.  Using eq.~(\ref{smat}) we can then extract the
phase shifts.  Our method for computing the phase shift actually allows us
to resolve the $2\pi$ ambiguity for each $\omega$ individually and has
other numerical advantages, which are elaborated in Ref.~\cite{method}.

For any value of $k$ we can obtain $\delta_{F}(k)$, so we can compute the
integral in eq.~(\ref{cas3}).  To find the bound state energies we solve
the eigenvalue problem numerically.  Levinson's theorem tells us how many
bound states to search for.  For a fixed background $\vec{\phi}$, the numerical
evaluation of eq.~(\ref{cas3}) can be done quickly and with high accuracy. 
This allows us to search over a class of $\vec{\phi}$'s for the
configuration with the lowest total energy.

\section{The Total Energy}

We are interested in calculating the total one loop effective 
energy of a static configuration $\vec{\phi}(x)$. We take $\vec{\phi}(x)$
to be specified by a short list of parameters $\{\zeta_i\}$. We 
measure energy in units of the fermion mass $m=Gv$ and use a 
dimensionless distance $\xi=mx$. In $1+1$ dimensions $\vec{\phi}(x)$
and $v$ are dimensionless.  We rescale $\vec{\phi}(x)$ by $v$ so that 
$\vec{\phi}(x)\to(1,0)$ as $|\xi|\to\infty$, and define dimensionless
couplings
\be
\tilde{\alpha}=\frac{\alpha}{G^2}
\qquad {\rm and} \qquad
\tilde{\lambda}=\frac{\lambda}{G^2}\, .
\label{workvar}
\ee
By this rescaling, using eq.~(\ref{lagb}) and eq.~(\ref{lag2}), we have
\be
\frac{E_{\rm cl}[\vec\phi\,]}{m}&=&
v^2\int_{-\infty}^\infty d\xi
\left( \frac{1}{2} \vec{\phi}\,'\cdot\vec{\phi}\,'+
\frac{\tilde{\lambda}}{8} [\vec{\phi}\cdot\vec{\phi}
-1+\frac{2\tilde{\alpha}}{\tilde{\lambda}}]^2
-\frac{\tilde{\lambda}}{2}
(\frac{\tilde{\alpha}}{\tilde{\lambda}})^2
-\tilde{\alpha}(\phi_1-1)\right)
\nonumber \\
&=&v^2{\cal E}_{\rm cl}(\tilde{\alpha},\tilde{\lambda},\{\zeta_{i}\})
\, ,
\label{ecl}
\ee
where prime denotes differentiation with respect to $\xi$. 

The fermion one loop contribution to the energy arises from eq.~(\ref{dirham}),
which with $\vec{\phi}$ measured in units of $v$ is
\be
H[\vec\phi\,] = m\left(i\sigma_1 \frac{d}{d\xi}
+\sigma_2\phi_1(\xi)+\sigma_3\phi_2(\xi)\right).
\label{dirhams}
\ee
We see that a single fermion makes a contribution proportional to $m$
and dependent on the variational parameters $\{\zeta_i\}$.  This means that
eq.~(\ref{cas3}) can be expressed as $m{\cal E}^F(\{\zeta_i\})$.  For $N_F$
flavors the one loop contribution is therefore $N_Fm{\cal
E}^F(\{\zeta_i\})$.

The boson one loop contribution comes from summing the square roots
of the eigenvalues of the operator
$-\frac{\partial^2}{\partial x^2}
+\frac{\partial^2 V}{\partial \phi_i\partial \phi_j}$.
Rescaling as before we find that the boson one loop energy can be
written as $m{\cal E}^B(\{\zeta_i\})$. Putting together the classical
energy and the one loop energies we get
\be
\frac{E_{\rm tot}[\vec\phi\,]}{N_F\, m}=
\frac{v^2}{N_F}
{\cal E}_{\rm cl}(\tilde{\alpha},\tilde{\lambda},\{\zeta_i\})
+{\cal E}^{F}(\{\zeta_i\})
+\frac{1}{N_F}{\cal E}^{B}(\tilde{\alpha},\tilde{\lambda},\{\zeta_i\})
+ \hbox{higher loops} \, .
\label{loopcount}
\ee
For $N_F$ large we can neglect the boson one loop contribution
relative to the fermion one loop contribution. Furthermore it can be
shown that $1/v^2$ counts boson loops. Taking both $N_F$ and $v^2$
large with the ratio fixed, we can neglect the higher loops entirely and
all but the single fermion loop in eq.~(\ref{loopcount}).  Therefore
we need only consider the contributions from ${\cal E}_{\rm cl}$ and
${\cal E}^F$.

\section{The Fermion Number}

A non--trivial background $\vec\phi(x)$ distorts the energy levels of the
Dirac Hamiltonian eq.~(\ref{dirham}), possibly introducing single particle
bound states  (with positive and negative energy). We identify the lowest
energy state of the system, the one with the all the negative energy levels
filled, as the vacuum.  If a level crosses zero as we locally
interpolate between $\vec\phi_{\rm cl}(x)$ and $\vec\phi(x)$, this
state will have non--zero fermion number.  In particular, if
$\vec{\phi}(x)$ circles $\vec{\phi}=(0,0)$ as $\vec{\phi}$ goes from
$(1,0)$ at  $x=-\infty$ to $(1,0)$ at $x=\infty$, then the vacuum
state will carry non--zero fermion number provided that the scale over
which $\vec\phi$ varies, $w$, is much larger than the fermion Compton
wavelength $1/m$ \cite{GoldWil,Farhi}.  In Ref.~\cite{tbaglevi} we
derive a formula for the fermion number of the vacuum, ${\cal Q}_{\rm
vac}$, in terms of the positive energy phase shifts at $k=0$ and the
number of positive energy bound states,
\be
{\cal Q}_{\rm vac}=N_F \left(\frac{1}{\pi}
\left[\delta_+(m)+\delta_-(m)\right]+\frac{1}{2}-n_{\omega>0} \right)
\label{qvac}
\ee
where $n_{\omega>0}$ is the number of bound states with positive
energy.  The configurations we look at loop at most once around
$\vec\phi=0$, so ${\cal Q}_{\rm vac}$ is either $0$ or $N_F$. We are
interested in states with fermion number $N_F$. If ${\cal Q}_{\rm
vac}=N_F$, then the state we want is the vacuum.  If ${\cal Q}_{\rm
vac}=0$, then we build the lowest energy state with fermion number $N_F$ by
filling the lowest positive energy level of~eq.~(\ref{dirham}) with $N_F$
fermions.  Therefore, if ${\cal Q}_{\rm vac} = 0$, ${\cal E}^F$ appearing
in eq.~(\ref{loopcount}) must be augmented by $\omega_1$ where $m\omega_1$
is the smallest positive eigenvalue of eq.~(\ref{dirhams}).

\section{Results}

We want to look for background configurations $\vec{\phi}$ that can produce
states with fermion number $N_F$, and whose total energy is below $N_Fm$.
From eq.~(\ref{loopcount}) with ${\cal E}^B$ neglected, we define
\be {\cal B}=\frac{v^2}{N_F} {\cal E}_{\rm cl} + {\cal E}^{F}-1\, ,
\label{etot}
\ee
which is the energy of the fermionic configuration minus the energy of
$N_F$ free fermions in units of $m N_F$.  For our numerical computations,
we take the {\it ansatz}
\be
\phi_1 +i \phi_2 & = &
1-R+R\, e^{i\Theta}
\quad {\rm where}\quad
\Theta=\pi(1+\tanh(\xi/w)) \, .
\label{var1}
\ee
The two variational parameters are $R$ and $w$. As $\xi$ goes
from $-\infty$ to $\infty$, $\vec{\phi}$ moves in a circle of radius
$R$ in the $(\phi_1,~\phi_2)$ plane, starting and ending at $(1,0)$.
The scale over which $\vec{\phi}$ varies is $w$.

\begin{figure}[t]
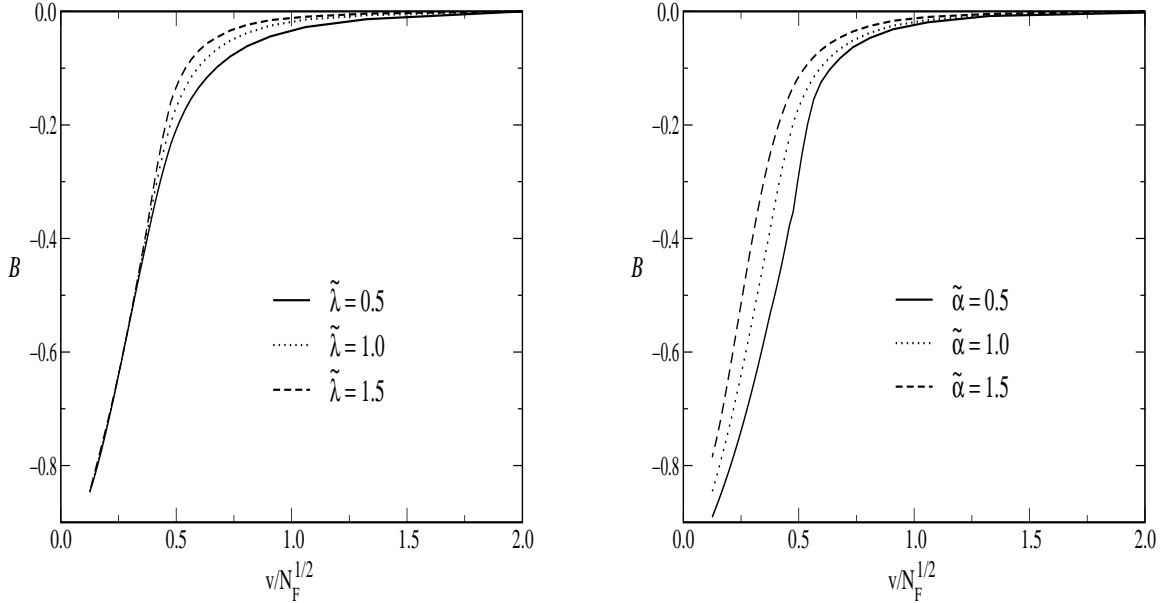

\centerline{
\epsfig{figure=lambda.eps,height=7.0cm,width=8.0cm,angle=270}
\hspace{1cm}
\epsfig{figure=alpha.eps,height=7.0cm,width=8.0cm,angle=270}}
\vskip0.5cm
\caption{\label{fig_1}{\sf ${\cal B}$ as a function of $v/\sqrt{N_F}$
for various values $\tilde{\lambda}$ with $\tilde{\alpha}=0.25$ (left
panel) and for various values $\tilde{\alpha}$ with
$\tilde{\lambda}=1.0$ (right panel).  ${\cal B}$ negative corresponds
to binding.}}
\end{figure}

For fixed $\tilde{\alpha}$ and $\tilde{\lambda}$ we vary $R$ and  $w$ until
we produce the configuration with the smallest  ${\cal B}$. The results are
shown in Fig.~\ref{fig_1}. We see that it is possible to find
a configuration whose total energy is below $N_Fm$.  Since we are minimizing
${\cal B}$ subject to the constraint that $\phi$ is of the form
eq.~(\ref{var1}), we know that the true minimum of ${\cal B}$ in the fermion
number $N_F$ sector also has an energy below $N_F m$.  This is the stable
soliton.

\begin{figure}[t]
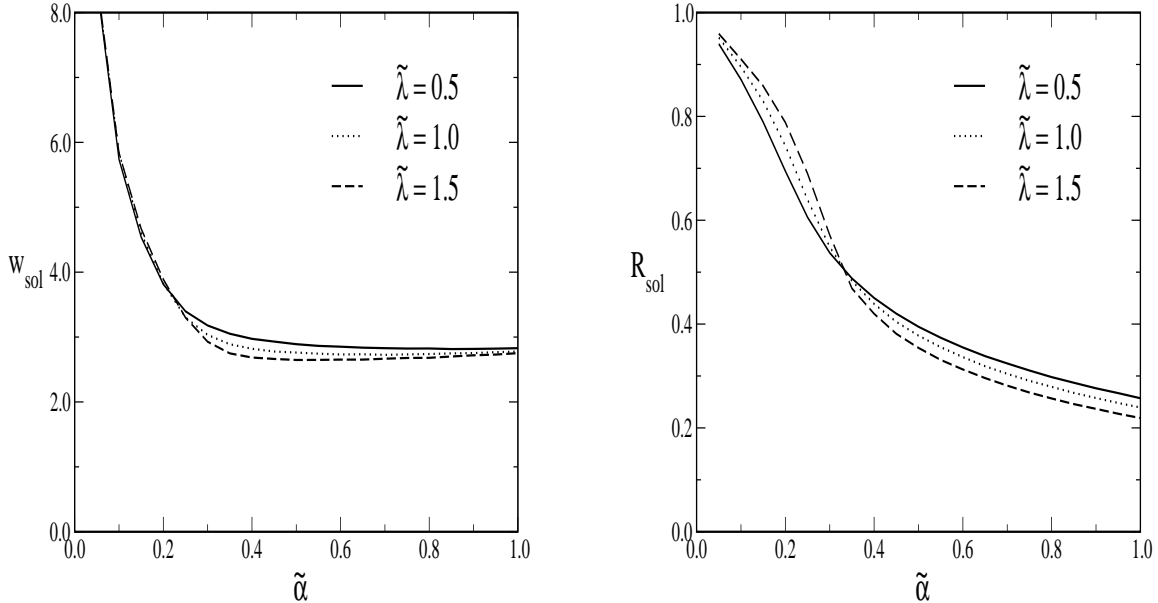

\centerline{
\epsfig{figure=width.eps,height=7.0cm,width=8.0cm,angle=270}
\hspace{1cm}
\epsfig{figure=radius.eps,height=7.0cm,width=8.0cm,angle=270}}
\vskip0.5cm
\caption{\label{fig_2}{\sf The width $w_{\rm sol}$ (left panel)
and the radius $R_{\rm sol}$ (right panel) of the
configurations (\protect\ref{var1}) that minimize the total energy
as a function of the explicit symmetry breaking $\tilde{\alpha}$.
Several values of the Higgs coupling constant $\tilde{\lambda}$
are considered and $v/\sqrt{N_F}=0.375$.}}
\end{figure}

In Fig.~\ref{fig_2} we show the width, $w_{\rm sol}$, and the radius,
$R_{\rm sol}$, for the minimum energy configuration as a function of
$\tilde{\alpha}$ for various values of $\tilde{\lambda}$.  Note that the
size of the soliton grows like $1/\sqrt{\tilde \alpha}$ as $\tilde{\alpha}$
goes to zero.  In that region, $R_{\rm sol}$ approaches 1, so the $\vec \phi$
configuration approaches the chiral circle.  In fact, the energy of the
fermion number $N_F$ soliton  goes to zero as $\tilde \alpha$ goes to
zero.  However for $\tilde \alpha$ very small the bosonic quantum
fluctuations restore the classically broken symmetry.  Thus we can not
trust our results for $\tilde \alpha$ very small and we do not believe that
this large and light soliton is a reliable consequence of this  model. 
For moderate values of $\tilde \alpha$, where the width of
the soliton is not controlled by $1/\sqrt{\tilde\alpha}$, we do trust
our results.  For the value of $v/\sqrt{N_F}$ shown in
Fig.~\ref{fig_2}, the model becomes trustworthy for $\tilde \alpha
\approx 0.3$.  For further discussion of this point, see
Ref.~\cite{longpaper}.

We have developed and applied a variational technique for
renormalizable quantum field theories through one loop order.  Because
we have applied unambiguous, standard perturbative renormalization
procedures, we have been able to hold the theory (i.e.  the
renormalized masses and coupling constants) fixed, while searching
over a variational ansatz.  Here we have used these methods to
demonstrate the existence of a stable fermionic soliton, stabilized by
quantum effects, in a model with no soliton at the classical level.
The result suggests that similar phenomena might persist in
3+1~dimensions, and no obstacles stand in the way of generalizing the 
method to that case.

\section*{Acknowledgments} We would like to thank S.~Bashinsky,
Y.~Frishman, J.~Goldstone, D.~Son, U.~Weise, X.~Wen, and F.~Wilczek
for helpful conversations, suggestions and references.  This work is
supported in part by funds provided by the U.S. Department of Energy
(D.O.E.) under cooperative research agreement \#DF-FC02-94ER40818 and
the Deutsche Forschungsgemeinschaft (DFG) under contract We 1254/3-1.

\end{document}